\documentclass[aps,prl,twocolumn,superscriptaddress,showpacs,letterpaper]{revtex4}

\usepackage{graphicx}
\usepackage{amsmath, amsthm, amssymb}

\usepackage{xspace}

\newcommand{\Bs}{\ensuremath{B^0_{s}}\xspace}
\newcommand{\BsH}{\ensuremath{B^0_{sH}}\xspace}
\newcommand{\BsL}{\ensuremath{B^0_{sL}}\xspace}
\newcommand{\Bd}{\ensuremath{B^{0}}\xspace}
\newcommand{\DG}{\ensuremath{\Delta\Gamma}\xspace}
\newcommand{\phiS}{\ensuremath{2\beta_s}\xspace}
\newcommand{\Jpsi}{\ensuremath{J\!/\!\psi}\xspace}
\newcommand{\BsDecay}{\ensuremath{B^0_s \to \Jpsi\,\phi}\xspace}
\newcommand{\BdDecay}{\ensuremath{B^0 \to \Jpsi\,K^*}\xspace}
\newcommand{\CP}{\textit{CP}\xspace}

\newcommand{\PaperTitle}{Measurement of Lifetime and Decay-Width Difference in \BsDecay Decays}

\begin{document}

\title{\PaperTitle}

\affiliation{Institute of Physics, Academia Sinica, Taipei, Taiwan 11529, Republic of China} 
\affiliation{Argonne National Laboratory, Argonne, Illinois 60439} 
\affiliation{Institut de Fisica d'Altes Energies, Universitat Autonoma de Barcelona, E-08193, Bellaterra (Barcelona), Spain} 
\affiliation{Baylor University, Waco, Texas  76798} 
\affiliation{Istituto Nazionale di Fisica Nucleare, University of Bologna, I-40127 Bologna, Italy} 
\affiliation{Brandeis University, Waltham, Massachusetts 02254} 
\affiliation{University of California, Davis, Davis, California  95616} 
\affiliation{University of California, Los Angeles, Los Angeles, California  90024} 
\affiliation{University of California, San Diego, La Jolla, California  92093} 
\affiliation{University of California, Santa Barbara, Santa Barbara, California 93106} 
\affiliation{Instituto de Fisica de Cantabria, CSIC-University of Cantabria, 39005 Santander, Spain} 
\affiliation{Carnegie Mellon University, Pittsburgh, PA  15213} 
\affiliation{Enrico Fermi Institute, University of Chicago, Chicago, Illinois 60637} 
\affiliation{Comenius University, 842 48 Bratislava, Slovakia; Institute of Experimental Physics, 040 01 Kosice, Slovakia} 
\affiliation{Joint Institute for Nuclear Research, RU-141980 Dubna, Russia} 
\affiliation{Duke University, Durham, North Carolina  27708} 
\affiliation{Fermi National Accelerator Laboratory, Batavia, Illinois 60510} 
\affiliation{University of Florida, Gainesville, Florida  32611} 
\affiliation{Laboratori Nazionali di Frascati, Istituto Nazionale di Fisica Nucleare, I-00044 Frascati, Italy} 
\affiliation{University of Geneva, CH-1211 Geneva 4, Switzerland} 
\affiliation{Glasgow University, Glasgow G12 8QQ, United Kingdom} 
\affiliation{Harvard University, Cambridge, Massachusetts 02138} 
\affiliation{Division of High Energy Physics, Department of Physics, University of Helsinki and Helsinki Institute of Physics, FIN-00014, Helsinki, Finland} 
\affiliation{University of Illinois, Urbana, Illinois 61801} 
\affiliation{The Johns Hopkins University, Baltimore, Maryland 21218} 
\affiliation{Institut f\"{u}r Experimentelle Kernphysik, Universit\"{a}t Karlsruhe, 76128 Karlsruhe, Germany} 
\affiliation{Center for High Energy Physics: Kyungpook National University, Daegu 702-701, Korea; Seoul National University, Seoul 151-742, Korea; SungKyunKwan University, Suwon 440-746, Korea; Korea Institute of Science and Technology Information, Daejeon, 305-806, Korea; Chonnam National University, Gwangju, 500-757, Korea} 
\affiliation{Ernest Orlando Lawrence Berkeley National Laboratory, Berkeley, California 94720} 
\affiliation{University of Liverpool, Liverpool L69 7ZE, United Kingdom} 
\affiliation{University College London, London WC1E 6BT, United Kingdom} 
\affiliation{Centro de Investigaciones Energeticas Medioambientales y Tecnologicas, E-28040 Madrid, Spain} 
\affiliation{Massachusetts Institute of Technology, Cambridge, Massachusetts  02139} 
\affiliation{Institute of Particle Physics: McGill University, Montr\'{e}al, Canada H3A~2T8; and University of Toronto, Toronto, Canada M5S~1A7} 
\affiliation{University of Michigan, Ann Arbor, Michigan 48109} 
\affiliation{Michigan State University, East Lansing, Michigan  48824} 
\affiliation{University of New Mexico, Albuquerque, New Mexico 87131} 
\affiliation{Northwestern University, Evanston, Illinois  60208} 
\affiliation{The Ohio State University, Columbus, Ohio  43210} 
\affiliation{Okayama University, Okayama 700-8530, Japan} 
\affiliation{Osaka City University, Osaka 588, Japan} 
\affiliation{University of Oxford, Oxford OX1 3RH, United Kingdom} 
\affiliation{University of Padova, Istituto Nazionale di Fisica Nucleare, Sezione di Padova-Trento, I-35131 Padova, Italy} 
\affiliation{LPNHE, Universite Pierre et Marie Curie/IN2P3-CNRS, UMR7585, Paris, F-75252 France} 
\affiliation{University of Pennsylvania, Philadelphia, Pennsylvania 19104} 
\affiliation{Istituto Nazionale di Fisica Nucleare Pisa, Universities of Pisa, Siena and Scuola Normale Superiore, I-56127 Pisa, Italy} 
\affiliation{University of Pittsburgh, Pittsburgh, Pennsylvania 15260} 
\affiliation{Purdue University, West Lafayette, Indiana 47907} 
\affiliation{University of Rochester, Rochester, New York 14627} 
\affiliation{The Rockefeller University, New York, New York 10021} 
\affiliation{Istituto Nazionale di Fisica Nucleare, Sezione di Roma 1, University of Rome ``La Sapienza," I-00185 Roma, Italy} 
\affiliation{Rutgers University, Piscataway, New Jersey 08855} 
\affiliation{Texas A\&M University, College Station, Texas 77843} 
\affiliation{Istituto Nazionale di Fisica Nucleare, University of Trieste/\ Udine, 34127 Trieste, Italy} 
\affiliation{University of Tsukuba, Tsukuba, Ibaraki 305, Japan} 
\affiliation{Tufts University, Medford, Massachusetts 02155} 
\affiliation{Waseda University, Tokyo 169, Japan} 
\affiliation{Wayne State University, Detroit, Michigan  48201} 
\affiliation{University of Wisconsin, Madison, Wisconsin 53706} 
\affiliation{Yale University, New Haven, Connecticut 06520} 
\author{T.~Aaltonen}
\affiliation{Division of High Energy Physics, Department of Physics, University of Helsinki and Helsinki Institute of Physics, FIN-00014, Helsinki, Finland}
\author{A.~Abulencia}
\affiliation{University of Illinois, Urbana, Illinois 61801}
\author{J.~Adelman}
\affiliation{Enrico Fermi Institute, University of Chicago, Chicago, Illinois 60637}
\author{T.~Akimoto}
\affiliation{University of Tsukuba, Tsukuba, Ibaraki 305, Japan}
\author{M.G.~Albrow}
\affiliation{Fermi National Accelerator Laboratory, Batavia, Illinois 60510}
\author{B.~\'{A}lvarez~Gonz\'{a}lez}
\affiliation{Instituto de Fisica de Cantabria, CSIC-University of Cantabria, 39005 Santander, Spain}
\author{S.~Amerio}
\affiliation{University of Padova, Istituto Nazionale di Fisica Nucleare, Sezione di Padova-Trento, I-35131 Padova, Italy}
\author{D.~Amidei}
\affiliation{University of Michigan, Ann Arbor, Michigan 48109}
\author{A.~Anastassov}
\affiliation{Rutgers University, Piscataway, New Jersey 08855}
\author{A.~Annovi}
\affiliation{Laboratori Nazionali di Frascati, Istituto Nazionale di Fisica Nucleare, I-00044 Frascati, Italy}
% ...
\collaboration{CDF Collaboration\footnote{With visitors from $^a$University of Athens, 15784 Athens, Greece, 
$^b$Chinese Academy of Sciences, Beijing 100864, China, 
$^c$University of Bristol, Bristol BS8 1TL, United Kingdom, 
$^d$University Libre de Bruxelles, B-1050 Brussels, Belgium, 
$^e$University of California Irvine, Irvine, CA  92697, 
$^f$University of California Santa Cruz, Santa Cruz, CA  95064, 
$^g$Cornell University, Ithaca, NY  14853, 
$^h$University of Cyprus, Nicosia CY-1678, Cyprus, 
$^i$University College Dublin, Dublin 4, Ireland, 
$^j$University of Edinburgh, Edinburgh EH9 3JZ, United Kingdom, 
$^k$University of Heidelberg, D-69120 Heidelberg, Germany, 
$^l$Universidad Iberoamericana, Mexico D.F., Mexico, 
$^m$University of Manchester, Manchester M13 9PL, England, 
$^n$Nagasaki Institute of Applied Science, Nagasaki, Japan, 
$^o$University de Oviedo, E-33007 Oviedo, Spain, 
$^p$Queen Mary, University of London, London, E1 4NS, England, 
$^q$Texas Tech University, Lubbock, TX  79409, 
$^r$IFIC(CSIC-Universitat de Valencia), 46071 Valencia, Spain, 
}}
\noaffiliation

\begin{abstract}
We measure the mean lifetime, $\tau=2/(\Gamma_L+\Gamma_H)$, and the 
decay-width difference, $\DG=\Gamma_L-\Gamma_H$, of the light and heavy mass eigenstates
of the \Bs meson, \BsL and \BsH, in \BsDecay decays using $1.7$ $\mathrm{fb^{-1}}$ 
of data collected with the CDF II detector 
at the Fermilab Tevatron $p\overline{p}$ collider.  
Assuming \CP conservation, a good approximation
for the \Bs system in the standard model,
we obtain $\DG = 0.076^{+0.059}_{-0.063}\mathrm{(stat.)} 
\pm 0.006\mathrm{(syst.)}$ $\mathrm{ps^{-1}}$ and
$\tau = 1.52 \pm 0.04\mathrm{(stat.)} \pm 0.02\mathrm{(syst.)}$ ps,
the most precise measurements to date.
Our constraints on the weak phase and \DG are consistent with \CP conservation.

\begin{center}
Dedicated to the memory of our dear friend and colleague, Michael P. Schmidt
\end{center}
\end{abstract}

\maketitle

In the standard model (SM), the mass and flavor eigenstates of the \Bs
meson differ. 
This gives rise to particle-antiparticle 
oscillations \cite{mixing}, which proceed in the SM through weak interaction 
processes, and whose phenomenology depends on the
Cabibbo-Kobayashi-Maskawa (CKM) quark mixing matrix.
The time ($t$) evolution of \Bs mesons is governed by the Schr\"odinger equation
\begin{equation*}
i\frac{d}{dt}\left( \left| B_s^0(t) \right> \atop \left| \bar{B}_s^0(t) \right> \right)
= \left( \mathbf{M} - \frac{i}{2}\mathbf{\Gamma} \right) \left( \left| B_s^0(t) \right> \atop \left| \bar{B}_s^0(t) \right> \right),
\end{equation*}
with mass matrix $\mathbf{M}$ and decay matrix $\mathbf{\Gamma}$.
The mass difference, $\Delta m=m_H-m_L \approx 2|M_{12}|$, between the heavy and light mass
eigenstates, \BsH and \BsL, determines the frequency 
of \Bs oscillations, a quantity precisely
measured in Ref.~\cite{Abulencia:2006ze}.
The mean lifetime, $\tau=2/(\Gamma_L + \Gamma_H)$, is expected to be
equal to the mean \Bd lifetime within $1\,\%$ \cite{lifetime}.
The decay-width difference, $\DG = \Gamma_L - \Gamma_H$, 
is predicted in the SM to be $0.096 \pm 0.039$ ps$^{-1}$ \cite{Lenz:2006hd}
and was first measured by the CDF Collaobarion \cite{Acosta:2004gt} 
and, recently, by the D0 Collaboration with higher precision \cite{Abazov:2007tx}. 
It depends on the \CP-violating weak phase difference
between the \Bs-$\bar{B}_s^0$ mixing amplitude and the amplitudes of the
subsequent \Bs and $\bar{B}_s^0$ decays to common final states, $\phi_s=\arg(-M_{12}/\Gamma_{12})$, 
via the relation $\DG = 2|\Gamma_{12}|\cos(\phi_s)$.
While the SM expectation, $\phi_s^\mathrm{SM}=4\times 10^{-3}$ \cite{Lenz:2006hd}, is
small, contributions from new physics processes to \Bs mixing
can lead to a significantly different value of the phase, $\phi_s = \phi_s^\mathrm{SM} + \phi_s^\mathrm{NP}$.
The same new physics contribution, $\phi_s^\mathrm{NP}$, would be present in the relative
phase between mixing and $b \to c\bar{c}s$ quark transitions, 
$\phiS = 2\beta_s^\mathrm{SM} - \phi_s^\mathrm{NP}$, in which the SM contribution is defined 
in terms of CKM matrix elements by
$\beta_s^\mathrm{SM}=\arg(-V_{ts}V_{tb}^*/V_{cs}V_{cb}^*) \approx 0.02$ \cite{Lenz:2006hd}.
Since both $\phi_s^\mathrm{SM}$ and $\beta_s^\mathrm{SM}$ are significantly
smaller than the current experimental resolution,
we can approximate $\phiS=-\phi_s^\mathrm{NP}=-\phi_s$.
Thus the measurement of a sizable value of \phiS inconsistent with zero would indicate new physics.
In case of a non-zero $|\Gamma_{12}|$,
an analysis of time-dependent decay rates of \Bs mesons to two vector mesons
becomes sensitive to the weak phase \phiS,
even without information on the \Bs flavor at production,
because of the interference between \CP eigenstates.

In this Letter, we present the measurement of the \Bs
meson mean lifetime $\tau$ and decay-width difference $\DG$ using
\BsDecay decays followed by $\Jpsi\rightarrow \mu^+\mu^-$ and
$\phi\rightarrow K^+K^-$ decays. 
Charge-conjugate modes are implied throughout this paper.
We also extract information about the weak phase \phiS.
The final state is a mixture of \CP-even and \CP-odd states
that are distinguished using the angular distributions of the decay products.
Since the \Bs is a pseudoscalar and the \Jpsi and $\phi$
are vector mesons, the orbital angular momentum
between the two decay products can have the magnitudes $\ell =0, 1,$ or 2.
The final state is \CP-even in S- and D-wave decays and \CP-odd in P-wave decays.
The angular distributions are expressed in terms of three angles, $\theta_T$, $\phi_T$, and $\psi_T$, 
defined in the transversity basis \cite{Dighe:1998vk}. 
The angles $\theta_T$ and $\phi_T$ are the
polar and azimuthal angles of the $\mu^+$ in the rest frame
of the \Jpsi, where the $x$-axis is defined by the momentum direction of the \Bs
and the $xy$-plane by the $\phi \to K^+ K^-$ decay plane
with a positive $y$ component of the $K^+$ momentum. 
The angle $\psi_T$ is the polar angle of the $K^+$ with respect 
to the opposite of the \Bs flight direction in the $\phi$ rest frame.

The data were collected by the CDF II 
detector at the Fermilab Tevatron $p\bar{p}$ collider between February 2002 and January
2007, and correspond to an integrated luminosity of
$1.7$ $\mathrm{fb^{-1}}$.  
The CDF II detector \cite{Acosta:2004yw} consists of a magnetic
spectrometer surrounded by electromagnetic and hadronic 
calorimeters and muon detectors.  
The tracking system is composed of a silicon 
micro-strip detector \cite{Hill:2004qb} surrounded by an open-cell
drift chamber (COT) \cite{Affolder:2003ep}.
We detect muons in planes of multiwire drift chambers and scintillators \cite{Ascoli:1987av}
in the pseudorapidity range $|\eta| \le 1.0$.
Charged particle identification 
is provided by the time-of-flight system \cite{TOF}, complemented by 
the ionization-energy-loss measurement in the COT ($dE/dx$).
Events with $\Jpsi \rightarrow \mu^+ \mu^-$
decays used in this analysis were recorded using a dimuon trigger, which required
two oppositely-charged COT tracks matched to muon chamber track segments with 
a dimuon mass between 2.7 and 4.0 GeV/$c^2$.

\begin{figure}[t]
  \centering
  \includegraphics[width=2.6in]{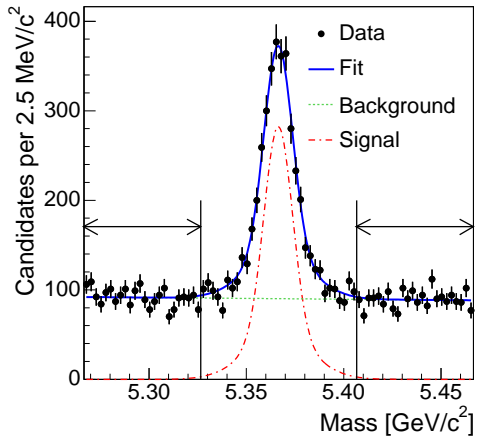}
  \caption{Invariant $\Jpsi\,\phi$ mass distribution with fit projection overlaid.
    The arrows indicate the sideband regions.}
  \label{fig:BsMass}
\end{figure}

%
In the offline analysis, \BsDecay decays are reconstructed following the
procedure described in Ref.~\cite{Acosta:2004gt}.
We train an artificial neural network (ANN) to
separate \Bs decays from the combinatorial background, which is the dominant one.
We model the signal with simulated events and use data from \Bs mass sidebands
(see Fig.~\ref{fig:BsMass}) to model the combinatorial background. 
The input variables to the ANN are kinematic quantities, vertex fit quality parameters,
and particle-identification information 
obtained from the muon system, the time-of-flight detector, and the $dE/dx$
measurements.
The requirement on the ANN output is selected by maximizing the significance
$S/\sqrt{S+B}$ on data where $S$ ($B$) is the number of signal
(background) events in a $\pm 20$ MeV/c$^2$ window around the \Bs mass peak position. 
The selected sample contains about $2500$ \BsDecay decays. 
The resulting mass distribution is shown in Fig.~\ref{fig:BsMass}.

To extract $\tau$ and \DG, we perform an unbinned maximum
likelihood fit with probability density functions (PDFs) depending on mass, lifetime, 
and transversity angles.

For the probability density functions (PDFs) of the background, we use empirical models with
floating fit parameters determined from the data. 
The background has a prompt component and a non-prompt component.
The mass PDF is parametrized by a straight-line function for each component.
The lifetime distribution is described by a delta function at $t=0$ for the prompt component,
a positive exponential for the long-lived non-prompt component, 
and a negative and positive exponential for mismeasured candidates. 
All lifetime components are convolved with a Gaussian to
account for the lifetime resolution
estimated on a candidate-by-candidate basis.
Because correlations among the three angles are negligible,
we factorize the angular PDF as a product of polynomials in
$\cos^2(\theta_T)$, $\cos(2\phi_T)$ and $\cos(\psi_T)$.
The angular distributions of prompt and non-prompt
background events agree within uncertainties,
and are chosen to be identical in the likelihood function.

For the signal, the mass distribution is described by the sum
of two Gaussians.
The lifetime and the angles $\vec{\rho}=(\cos(\theta_T),\phi_T,\cos(\psi_T))$
 are correlated for
\Bs signal events.
The lifetime-angular distribution without acceptance effects is given by
\begin{eqnarray}
\frac{d^4P(\vec{\rho},t)}{d\vec{\rho}dt} &\propto& 
 |A_0|^2f_1(\vec{\rho}){\cal T}_+ +
|A_{||}|^2f_2(\vec{\rho}){\cal T}_+\nonumber \\
 &+& |A_{\perp}|^2f_3(\vec{\rho}){\cal T}_- +
|A_0||A_{||}|f_5(\vec{\rho}) \cos(\delta_{||}){\cal
T}_+\nonumber \\
 &+& |A_{||}||A_\perp|f_4(\vec{\rho})
\cos(\delta_{\perp}-\delta_{||}) \nonumber \\
&&\sin(\phiS)
(e^{-\Gamma_Ht}-e^{-\Gamma_Lt})/2 \nonumber \\
 &+& |A_{0}||A_\perp|f_6(\vec{\rho})
\cos(\delta_{\perp}) \nonumber \\
&&\sin(\phiS) (e^{-\Gamma_Ht}-e^{-\Gamma_Lt})/2,
\label{eq:timedepangles}
\end{eqnarray}
where 
\begin{eqnarray*}
{\cal T}_\pm &=& \left[(1\pm\cos(\phiS))e^{-\Gamma_Lt}+(1\mp\cos(\phiS))e^{-\Gamma_Ht}\right]/2, \\
 f_1(\vec{\rho})&=& 2\cos^2(\psi_T)(1-\sin^2(\theta_T) \cos^2(\phi_T)), \\
 f_2(\vec{\rho})&=& \sin^2(\psi_T)(1-\sin^2(\theta_T) \sin^2(\phi_T)), \\
 f_3(\vec{\rho})&=& \sin^2(\psi_T) \sin^2(\theta_T), \\
 f_4(\vec{\rho})&=& -\sin^2(\psi_T) \sin(2\theta_T) \sin(\phi_T),
\\
 f_5(\vec{\rho})&=& \sin(2\psi_T) \sin^2(\theta_T) \sin(2\phi_T) / \sqrt{2}, \\
 f_6(\vec{\rho})&=& \sin(2\psi_T) \sin(2\theta_T) \cos(\phi_T) / \sqrt{2}.
\end{eqnarray*}
The quantities $A_0$, $A_\perp$ and $A_{||}$ are the linear polarization amplitudes at $t=0$, and
$\delta_{\perp}$ and $\delta_{||}$ are the strong phases of $A_\perp$ and $A_{||}$ 
relative to $A_0$, respectively. 

The lifetime-angle distribution is invariant under
each of the two transformations
$( \phiS \rightarrow -\phiS,\ 
\delta_\perp \rightarrow \delta_\perp + \pi )$ 
and
$( \Delta\Gamma \rightarrow -\Delta\Gamma,\ 
\phiS \rightarrow \phiS + \pi )$.
Because of this four-fold ambiguity, this measurement is
insensitive to the sign of both \phiS and \DG.

The signal lifetime terms are convolved
with the same Gaussian resolution function used for the background,
which employs the candidate-by-candidate lifetime uncertainty.
To account for different distributions of the lifetime uncertainty
between signal and background,
their PDFs are included in the likelihood.
These PDFs are derived from sideband-subtracted signal events and
from sideband events, respectively.

The angular distribution of \Bs decays described in 
Eq.~(\ref{eq:timedepangles}) is modified by detector acceptance as well as
trigger and selection efficiencies.
This effect is taken into account with an
acceptance function, $\epsilon(\vec{\rho})$, derived
from simulated \BsDecay decays.
The factor $\epsilon(\vec{\rho})$ is described by a three-dimensional histogram with $20$
bins in each of the angles.

We consider possible systematic uncertainties due to the signal mass model,
the lifetime resolution model,
the $\mathcal{O}(3\,\%)$ contamination by \BdDecay decays 
misreconstructed and selected as \Bs candidates
not included in the background model,
the acceptance description, 
the silicon detector alignment,
and the model for the angular distribution of the background.
The largest systematic uncertainty for \DG is caused by
\Bd mesons reconstructed as \Bs mesons.
The largest contributions to the systematic uncertainty on
$\tau$ are the lifetime resolution model
and the silicon detector alignment.
The dominant source of systematic uncertainties on the amplitudes is
the angular background model.

\begin{figure}[t]
        \centering
        \includegraphics[width=2.6in]{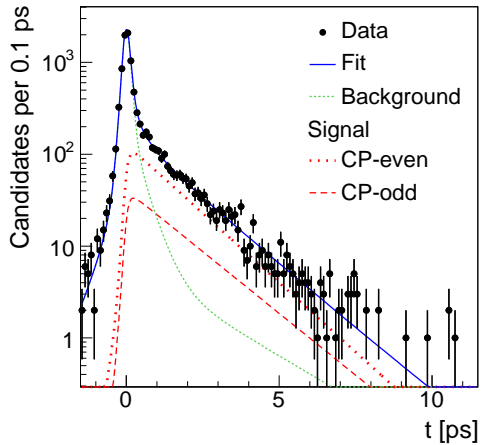}
        \caption{Lifetime distribution with fit projection overlaid.}
        \label{fig:BsLifetime}
\end{figure}
\begin{figure}[t]
  \centering
  \includegraphics[width=2.6in]{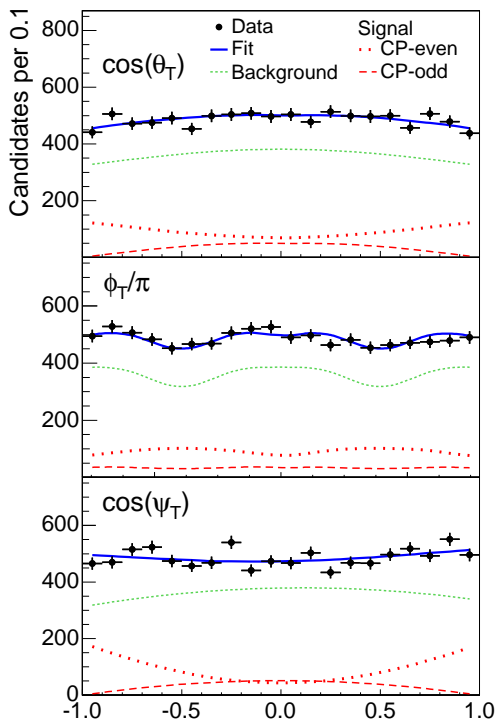}
  \caption{Angular distributions with fit projection overlaid.}
  \label{fig:angles}
\end{figure}

Under the assumption of \CP conservation ($\phiS=0$),
we obtain
\begin{eqnarray*}
  \tau            &=& 1.52 \pm 0.04   \pm 0.02\mbox{ ps}, \\ 
  \DG    &=& 0.076 ^{+0.059}_{-0.063} \pm 0.006 \mbox{ ps$^{-1}$}, \\
  |A_0|^2         &=& 0.531 \pm 0.020 \pm 0.007, \\ 
  |A_\perp|^2     &=& 0.239 \pm 0.029 \pm 0.011, \\ 
  |A_{||}|^2      &=& 0.230 \pm 0.026 \pm 0.009.
\end{eqnarray*}
The first uncertainties are statistical, the second ones systematic.
We do not quote a point-estimate of the strong phase $\delta_{||}$
because its likelihood profile is non-parabolic due to a symmetry point at $\delta_{||}=\pi$
which makes an uncertainty estimate unreliable.
This analysis is insensitive to the second strong phase $\delta_\perp$ if $\phiS=0$ 
(see Eq.(\ref{eq:timedepangles})).
The invariant mass, proper decay time, and angular distributions 
with fit projections overlaid are shown in Figs.~\ref{fig:BsMass} to \ref{fig:angles}.
The measured mean lifetime is compatible with the \Bd lifetime \cite{PDG2006} as predicted by theory \cite{lifetime}.
The measured amplitudes are consistent with the ones observed in \BdDecay decays \cite{BdAnalyses}
as expected under the assumption of SU(3) flavor symmetry.

For the constraints on the \CP-violating phase, we construct a $90$ $(95)\,\%$ confidence level region in
the \phiS--\DG plane using the likelihood-ratio ordering of Feldman and Cousins \cite{Feldman:1997qc}.
We choose this method instead of a point estimate, because it is not affected by the bias 
we observe in simulated experiments. 
The bias is of the order of the statistical uncertainty for input values of \DG or \phiS close to zero, 
which are near to the SM expectation. 
The bias can be understood from Eq.~(\ref{eq:timedepangles}).
If \phiS approaches zero, the two terms proportional to $\sin(\phiS)$
vanish, and this analysis becomes insensitive to $\delta_\perp$.
The same effective loss of degrees of freedom in the fit occurs when
\DG approaches zero and multiple degenerate solutions for \phiS and $\delta_\perp$ exist.

To obtain the likelihood ratio distribution for given values of \DG and \phiS,
we use experiments simulated with values for all other parameters 
determined by a fit to data \cite{PlugIn}.
We checked that alternate choices of these values do not
affect the coverage properties of our algorithm.
Systematic uncertainties are not included in the algorithm, 
since they are all negligible.

\begin{figure}[t]
        \centering
        \includegraphics[width=2.6in]{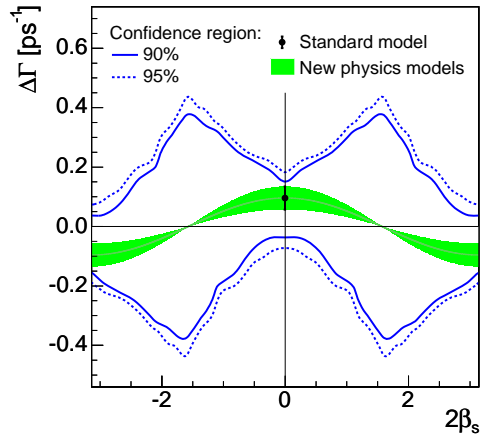}
        \caption{Regions at the $90\,\%$ and $95\,\%$ confidence level in the \phiS--\DG plane 
          compared with the SM prediction and the region allowed in new physics models
          given by $\DG = 2|\Gamma_{12}|\cos(\phi_s)$ \cite{Lenz:2006hd}.}
        \label{fig:confidenceRegion}
\end{figure}

The resulting confidence region is shown in Fig.~\ref{fig:confidenceRegion}. 
Since both \Bs mass eigenstates have the same angular distribution at $\phiS=\pm\pi/2$,
the sensitivity on \DG decreases towards this value.
For the SM expectation ($\DG \approx 0.1$ ps$^{-1}$ and $\phiS \approx 0$), we find 
the probability to get an equal or greater
likelihood ratio than the one observed in data to be $p=22\,\%$,
corresponding to an agreement at 1.2 Gaussian standard deviations.

In summary, we report the measurement of the mean lifetime,
the width difference, and the amplitudes in \BsDecay decays assuming \CP conservation. 
This measurement improves the precision of the current best measurement 
\cite{Abazov:2007tx} by 30-50$\,\%$.
It is in good agreement with previous results and the SM expectation. 
In addition we derive constraints on \DG and the \CP-violating phase \phiS.
Our data are consistent with the SM expectation of $\phiS\approx 0$, 
but sizeable values allowed within new physics
models cannot be ruled out. 

%
%
We thank U. Nierste for useful discussions.
We thank the Fermilab staff and the technical staffs of the participating 
institutions for their vital contributions. This work was supported by  
the U.S. Department of Energy and National Science Foundation; the 
Italian Istituto Nazionale di Fisica Nucleare; the Ministry of Education, 
Culture, Sports, Science and Technology of Japan; the Natural Sciences 
and Engineering Research Council of Canada; the National Science Council 
of the Republic of China; the Swiss National Science Foundation; the A.P. 
Sloan Foundation; the Bundesministerium f\"ur Bildung und Forschung, 
Germany; the Korean Science and Engineering Foundation and the Korean 
Research Foundation; the Science and Technology Facilities Council and the Royal 
Society, UK; the Institut National de Physique Nucleaire et  
Physique des Particules/CNRS; the Russian Foundation for Basic Research; 
the Comisi\'on Interministerial de Ciencia y Tecnolog\'{\i}a, Spain;
the European Community's Human Potential Programme; the
Slovak R\&D Agency; and 
the Academy of Finland. 
%




\begin{thebibliography}{99}

\bibitem{mixing}
C.~Gay, Annu.~Rev.~Nucl.~Part.~Sci.~\textbf{50}, 577 (2000).
\bibitem{Abulencia:2006ze}
A.~Abulencia \textit{et~al.} (CDF Collaboration),  
  Phys.~Rev.~Lett.~\textbf{97}, 242003 (2006).
\bibitem{lifetime}
  C.~Tarantino,
  Eur.\ Phys.\ J.\  C {\bf 33} (2004) S895
\bibitem{Lenz:2006hd}
A.~Lenz and U.~Nierste, J.~High Energy Phys.~\textbf{0706}, 072
(2007).
\bibitem{Acosta:2004gt}
D.~Acosta \textit{et~al.} (CDF Collaboration), 
Phys.~Rev.~Lett.~\textbf{94}, 101803 (2005).
\bibitem{Abazov:2007tx}
V.~M.~Abazov \textit{et~al.} (D0 Collaboration), 
Phys.~Rev.~Lett.~\textbf{98}, 121801 (2007).
\bibitem{Dighe:1998vk}
  A.~S.~Dighe, I.~Dunietz, and R.~Fleischer,
  Eur.\ Phys.\ J.\  C {\bf 6} 647 (1999);
A.~S.~Dighe \textit{et~al.}, 
  Phys.\ Lett.\  B {\bf 369} 144 (1996).
\bibitem{Acosta:2004yw}
D.~Acosta \textit{et~al.} (CDF Collaboration),
Phys.~Rev.~D \textbf{71}, 032001 (2005).
\bibitem{Hill:2004qb}
C.~S.~Hill, 
Nucl.~Instrum.~Methods A \textbf{530}, 1 (2004);
A.~Affolder \textit{et~al.},
\textit{ibid.} \textbf{453}, 84 (2000);
A.~Sill,
\textit{ibid.} \textbf{447}, 1 (2000).
\bibitem{Affolder:2003ep}
T.~Affolder \textit{et~al.},
Nucl.~Instrum.~Methods A \textbf{526}, 249 (2004).
\bibitem{Ascoli:1987av}
G.~Ascoli \textit{et~al.}, 
Nucl.~Instrum.~Methods A \textbf{268}, 33 (1988).
\bibitem{TOF} D.~Acosta \textit{et~al.}, 
Nucl.~Instrum.~Methods A \textbf{518}, 605 (2004).
\bibitem{PDG2006} 
   W.-M.~Yao {\it et al.}  (Particle Data Group),
   J.~Phys.~G {\bf 33}, 1 (2006).
\bibitem{BdAnalyses}
  B.~Aubert {\it et al.} ({\mbox{\sl B\hspace{-0.4em} {\small\sl A}\hspace{-0.37em} \sl B\hspace{-0.4em} {\small\sl A\hspace{-0.02em}R}}} Collaboration),
  Phys.~Rev.~D {\bf 76}, 031102 (2007);
  R.~Itoh {\it et al.}  (Belle Collaboration),
  Phys.~Rev.~Lett.  {\bf 95}, 091601 (2005).
\bibitem{Feldman:1997qc}
G.~J.~Feldman and R.~D.~Cousins, Phys.~Rev.~D
\textbf{57}, 3873 (1998).
\bibitem{PlugIn}
A.~C.~Davison and D.~V.~Hinkley, {\it Bootstrap methods and their applications}
(Cambridge University Press, 1997).


\end{thebibliography}
\end{document}